\documentclass{ws-procs975x65-vaso}
\usepackage{hyperref}
\usepackage{xspace}  

\def\beq{\begin{equation}}
\def\eeq{\end{equation}}
\newcommand{\met}{\ensuremath{E_{\mathrm T}^\text{miss}}\xspace}

\DeclareGraphicsExtensions{.pdf,.png,.jpg} 
\graphicspath{{./}{./figs/}}
\begin{document}

\mark{{V.~A.~Mitsou}{Dark matter: experimental and observational status}}

\vspace*{-1.4cm}
\begin{flushright}
IFIC/19-17
\end{flushright}

\title{Dark matter: experimental and observational status}
\author{Vasiliki~A.~Mitsou}

\address{Instituto de F\'isica Corpuscular (IFIC), CSIC -- Universitat de Val\`encia, \\
C/ Catedr\'atico Jos\'e Beltr\'an 2, E-46980 Paterna (Valencia), Spain\\
E-mail: vasiliki.mitsou@ific.uv.es\\
webific.ific.uv.es/web}

\begin{abstract}
This brief review covers recent results on searches for dark matter in collider experiments, as well as from direct and indirect detection observatories. It focuses on generic searches for dark matter signatures at the LHC, e.g.\ mono-$X$, dijets, etc. Recently observed astrophysical signals that may provide hints of dark matter are also discussed.
\end{abstract}

\keywords{Dark matter; Direct detection; Indirect detection; LHC.}

\bodymatter

%%%%%%%%%%%%%%%%% now a standard article style for the most part
%%%%%%%%%%%%%%%%%%%%%%%%%%%%%%%%%%%%%%%%%%%%%%%%%%%%
\section{Introduction}\label{sc:intro}

Dark matter (DM) still remains one of the most puzzling and fascinating mysteries in Fundamental Physics nowadays. The quest for unveiling its nature encompasses Cosmology, Astroparticle and Particle Physics.  Observations over the past decades --- obtained by combining a variety of astrophysical data, such as type-Ia supernovae, cosmic microwave background (CMB), baryon oscillations and weak lensing data --- indicate that most of our Universe energy budget consists of unknown entities: $\sim\!27\%$ is dark matter and $\sim\!68\%$ is dark energy,\cite{Aghanim:2018eyx} a form of ground-state energy. 

%Dark matter existence is inferred from gravitational effects on visible matter, but is undetectable by emitted or scattered electromagnetic radiation. The most precise measurement comes from CMB anisotropies. Experiments at high-energy physics colliders are expected to shed light to its nature and determine its properties. 

%%%%%%%%%%%%%%%%%%%%%%%%%%%%%%%%%%%%%%%%%%%%%%%%%%%%
\section{DM Candidates}\label{sc:candidates}

Weakly interacting massive particles (WIMPs) are the leading class of candidates for cold DM (CDM). They are thermal relics from Big Bang and the measured relic density implies that the DM annihilation cross section is of the same order as the one characterising the weak interaction scale, constituting the so-called ``WIMP miracle''. Specific theoretical models may provide naturally a DM particle, such as supersymmetry (SUSY), extra dimensions and little Higgs models. 

Other non-WIMP possibilities to explain the DM observations are superWIMPs (gravitinos, axinos), axions, (sterile) neutrinos, fuzzy CDM , Q-balls, WIMPzillas and macroscopic objects, such as primordial black holes.\cite{Kusenko:2013saa} New paradigms are continuously being proposed, partly as a result of interplay with observations. For instance, self-interacting DM has been proposed to ameliorate observed tensions between $N$-body simulations of collisionless CDM and astrophysical observations on galactic scales: cusp-vs-core problem, too-big-to-fail problem, missing-satellite problem, diversity problem.\cite{Spergel:1999mh,Mavromatos:2016vbj,nick} 

%%%%%%%%%%%%%%%%%%%%%%%%%%%%%%%%%%%%%%%%%%%%%%%%%%%%
\section{Direct Detection}\label{sc:direct}

Direct detection (DD) of DM involves the observation of elastic scattering of WIMP off nuclei. It is sensitive to the WIMP mass $m_\chi$ and the cross section $\sigma_{\chi\text{--nucleon}}$. A low-energy threshold in the WIMP-induced recoils is required for an efficient DD experiment, as well as reliable shielding of the detector from radioactive sources and cosmic backgrounds. The background suppression relies on the ability to distinguish nuclear recoils against other possible processes, such as electrons and $\alpha$-particles. Lastly, stable detector operation during annual and diurnal modulation is required.

The use of different targets safeguards against nuclei-related systematic uncertainties. Liquid noble gases, such as Xenon and Argon, offer sensitivity over the widest WIMP-mass range from 5~GeV  to 1~TeV (Darkside, DARWIN, DEAP3600, LUX, LZ, Panda-X, XENON). The (oldest) technology of cryogenic crystals presents new innovations and covers $m_\chi \sim 1-10$~GeV (CRESST, EDELWEISS, SuperCDMS). Alternative targets with unique properties include NaI crystals and bubble chambers (ANAIS, COSINE, DAMA/LIBRA, SABRE, PICO). A recent review of DD concepts and status is given in Ref.~\refcite{Schumann:2019eaa}.

No signal of dark matter in direct detection has been observed so far. The only persisting ``anomaly'' over several years is the annually modulating signal   observed by DAMA/LIBRA, a massive array of low-background NaI(Tl) crystals installed in the Gran Sasso underground laboratory.  This modulation has been seen in various phases of the detector and its period and phase are consistent with the expectation from the standard DM halo model.\cite{Bernabei:2013xsa}

As the sensitivity in cross section lowers, it reaches the so-called ``neutrino floor'', i.e.\ where the (irreducible) neutrino-flux background becomes dominant. Some DM models, e.g.\ SUSY or Composite Higgs\cite{Yepes:2018zkk} models, are still viable in regions of the $\sigma_{\chi\text{--nucleon}}$-vs.-$m_\chi$ plane. Some caveats apply to the DD results. The astrophysical uncertainties on the local DM density and DM velocity distribution are typically large. On the particle physics side, the way DM interacts with the detector is only partially known, while other nuclear-physics uncertainties may be considerable.

%%%%%%%%%%%%%%%%%%%%%%%%%%%%%%%%%%%%%%%%%%%%%%%%%%%%
\section{Indirect Detection}\label{sc:indirect}

In the indirect detection (ID) of DM, the focus is on the DM-particle decay or annihilation products in the galactic centre, dwarf galaxies, etc. Several probes are used as messengers, such as neutrinos, photons, antiprotons, and positrons. This class of observations may distinguish among different WIMP candidates: neutralinos, KK states, etc. Detection and analysis techniques used in ID and results are detailed in Ref.~\refcite{Slatyer:2017sev}. Several hints of DM annihilation have been observed in data over the years frequently being attributed to astrophysical origin. Here, we outline two recent observations: excess of GeV gamma rays and the 21-cm signal line.

\begin{figure}[htb]
\begin{minipage}{0.44\linewidth}
\includegraphics[width=\linewidth]{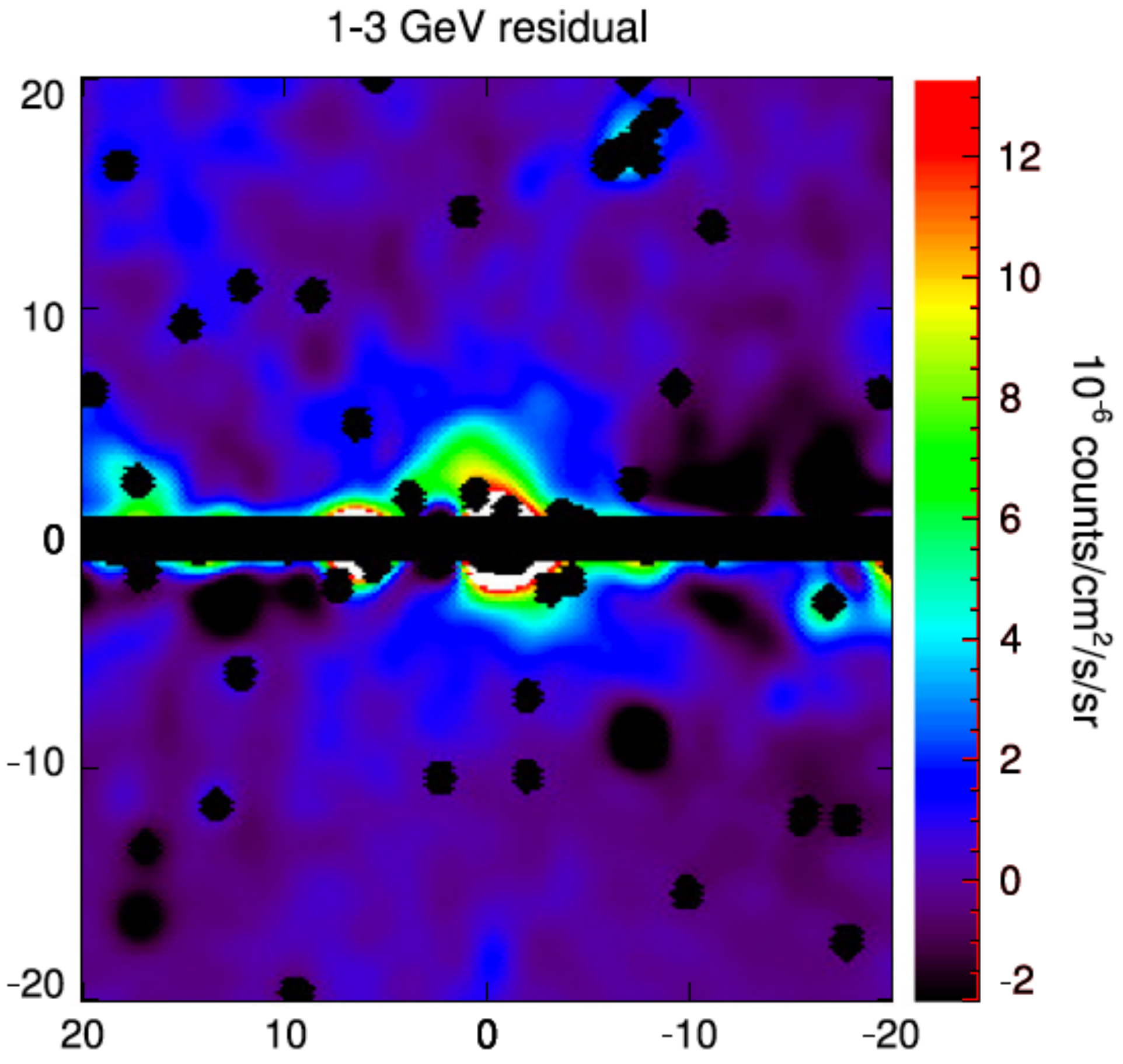}
\caption{\label{fg:gev-gamma-excess}Gamma-ray sky map after subtracting the point source model and the best-fit Galactic diffuse model, Fermi Bubbles, and isotropic templates. From Ref.~\protect\refcite{Daylan:2014rsa}. }
\end{minipage}\hspace{0.05\linewidth}%
\begin{minipage}{0.5\linewidth}
\includegraphics[width=\linewidth]{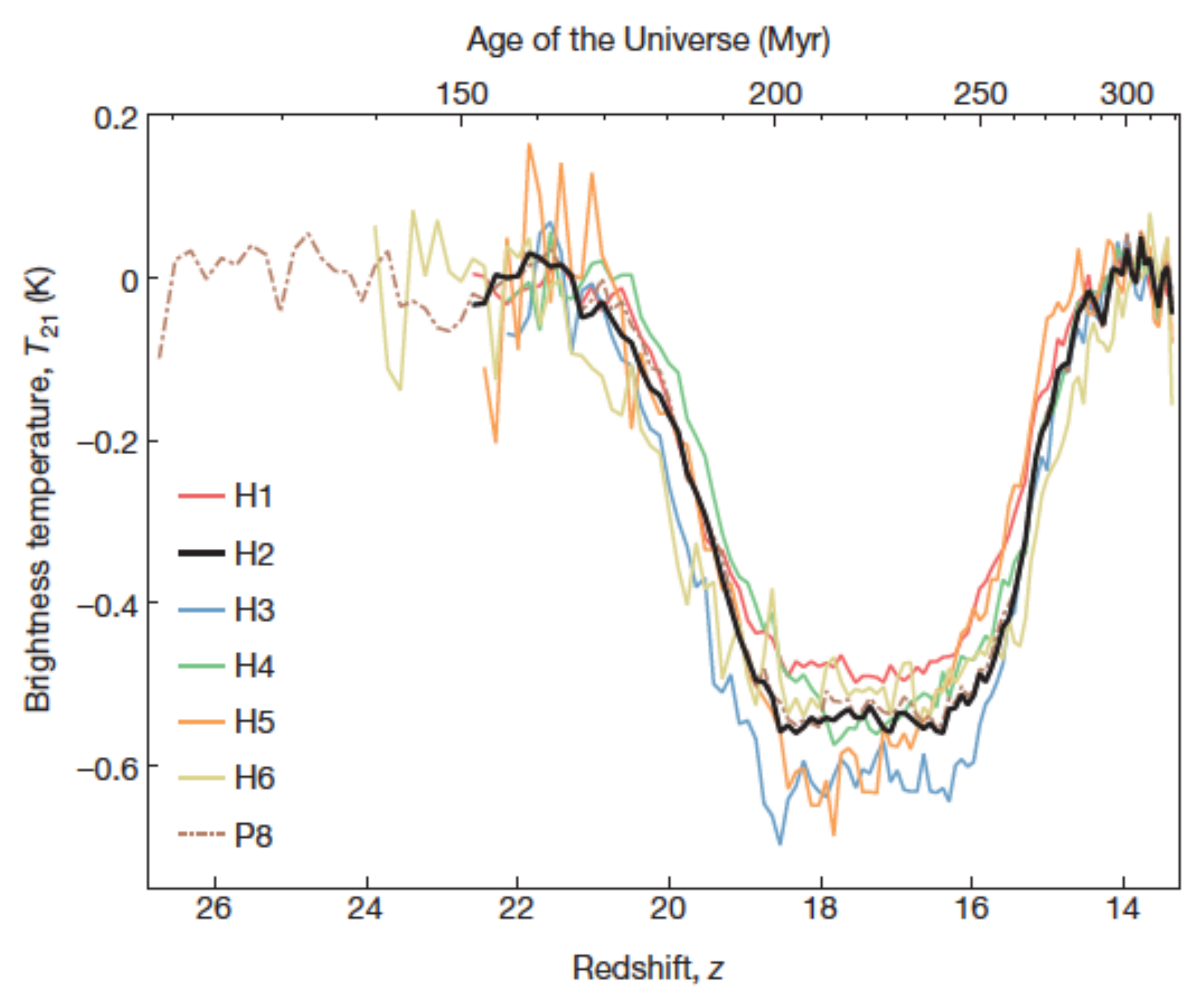}
\caption{\label{fg:edges}Best-fitting 21-cm absorption profiles for different hardware cases of the EDGES experiment. The thick black line is the model fit with the highest signal-to-noise ratio. From Ref.~\protect\refcite{Bowman:2018yin}. }
\end{minipage} 
\end{figure}

A spatially extended excess of $\sim1-3$~GeV $\gamma$ rays from the region surrounding the Galactic Center has been identified, consistent with the emission expected from annihilating dark matter. High resolution $\gamma$-ray maps, such as the one shown in Fig.~\ref{fg:gev-gamma-excess}, render the excess robust and highly statistically significant, with a spectrum, angular distribution, and overall normalisation that is in good agreement with that predicted by simple annihilating $36-51$~GeV DM particle annihilating to $b\bar{b}$.\cite{Daylan:2014rsa} 

The Experiment to Detect the Global EoR Signature (EDGES) is a radio-telescope for detection of hydrogen signatures from the Epoch of Reionisation (EoR), soon after the formation of the first stars and galaxies. The Collaboration reported the observation of an unexpectedly deep absorption in the radio background at 78~MHz, shown in Fig.~\ref{fg:edges}, based on their low-band instruments and interpreted it as a redshifted 21-cm line.\cite{Bowman:2018yin} It is unlikely that radiation from stars and stellar remnants to account for this discrepancy. Cooling of gas as a result of interactions between dark matter and baryons seems to explain the observed amplitude. 

%%%%%%%%%%%%%%%%%%%%%%%%%%%%%%%%%%%%%%%%%%%%%%%%%%%%
\section{Searches at the LHC}\label{sc:lhc}

The ATLAS and CMS experiments have embarked upon searches for signals of DM produced at the LHC early on during data taking.~\cite{Mitsou:2013rwa} Theoretical models, such as supersymmetry~\cite{Mitsou:2017mrq} or theories with extra-dimensions~\cite{Mitsou:2017ssh} provide a natural DM candidate, hence searches targeting these models implicitly cover dark matter, too. Since the candidate is only weakly interacting with matter, the common feature among these analyses is the requirement for large missing momentum \met. These models offer definite predictions, however they are characterised by a large number of parameters and this approach is profoundly model dependent. 

The first generic approach towards DM searches involved the deployment of effective field theories (EFTs), extensively used in Run~1 (2010--2012). In this method, the interaction between DM and Standard Model (SM) particles is described by effective operators. It is clearly less model independent, yet it is only valid for low-momentum transfers  $Q^2 \ll M_\text{med}^2$, where $M_\text{med}$ is the mediator mass.

A third approach extensively followed in Run~2 (2015--2018) assumes simplified topologies where the DM and SM particles interact via a mediator ($Z', H$).\cite{Boveia:2016mrp} Its advantage is that it covers features of a whole class of models and remains valid at high energies. In addition, it is described by a relatively small number of free parameters, namely the mediator and DM masses $M_\text{med}$ and $m_\chi$ and the mediator couplings to SM and DM particles $g_\text{SM}$ and $g_\text{DM}$, respectively.

Concerning the final states, DM cannot be directly observed at the LHC, however DM-pair production can be detected via the presence of an imbalance in measured transverse momenta of visible particles. The DM pair may only give rise to large \met, if it recoils to an energetic particle $X$, hence the so-called mono-$X$ searches, where $X$ is a jet,\cite{Sirunyan:2017jix} a photon,\cite{Aaboud:2017dor} a top, a (hadronically\cite{Sirunyan:2017jix} or leptonically~\cite{Sirunyan:2017qfc} decaying) $W/Z$ boson. When $X=H (\to b\bar{b}, \gamma\gamma, \tau\tau)$ coupled to DM through a BSM effective vertex, it provides a direct probe of the DM--SM coupling. Additionally, the associated production of DM with $t\bar{t}$\cite{Aaboud:2017aeu,Aaboud:2017rzf} and $b\bar{b}$\cite{Aaboud:2017rzf} pairs is considered. 

In addition to the dedicated analyses for DM, constraints can be extracted from recasting searches for heavy resonances. For an assumed $Z'$ mediator, the final states can be two quarks (``dijet''),\cite{Sirunyan:2017nvi,Sirunyan:2018xlo} two Higgs bosons ($\to b\bar{b}, \tau\tau$), $H\gamma \to b\bar{b}\gamma$. Such searches are only sensitive to the mediator--SM particle coupling, e.g.\ the mediator--quark coupling in the case of dijet resonances. A wide range of mediator masses are covered with these analyses with/without jet/$\gamma/W$ initial state radiation.

\begin{figure}[htb]
\begin{center}
\includegraphics[width=0.95\linewidth]{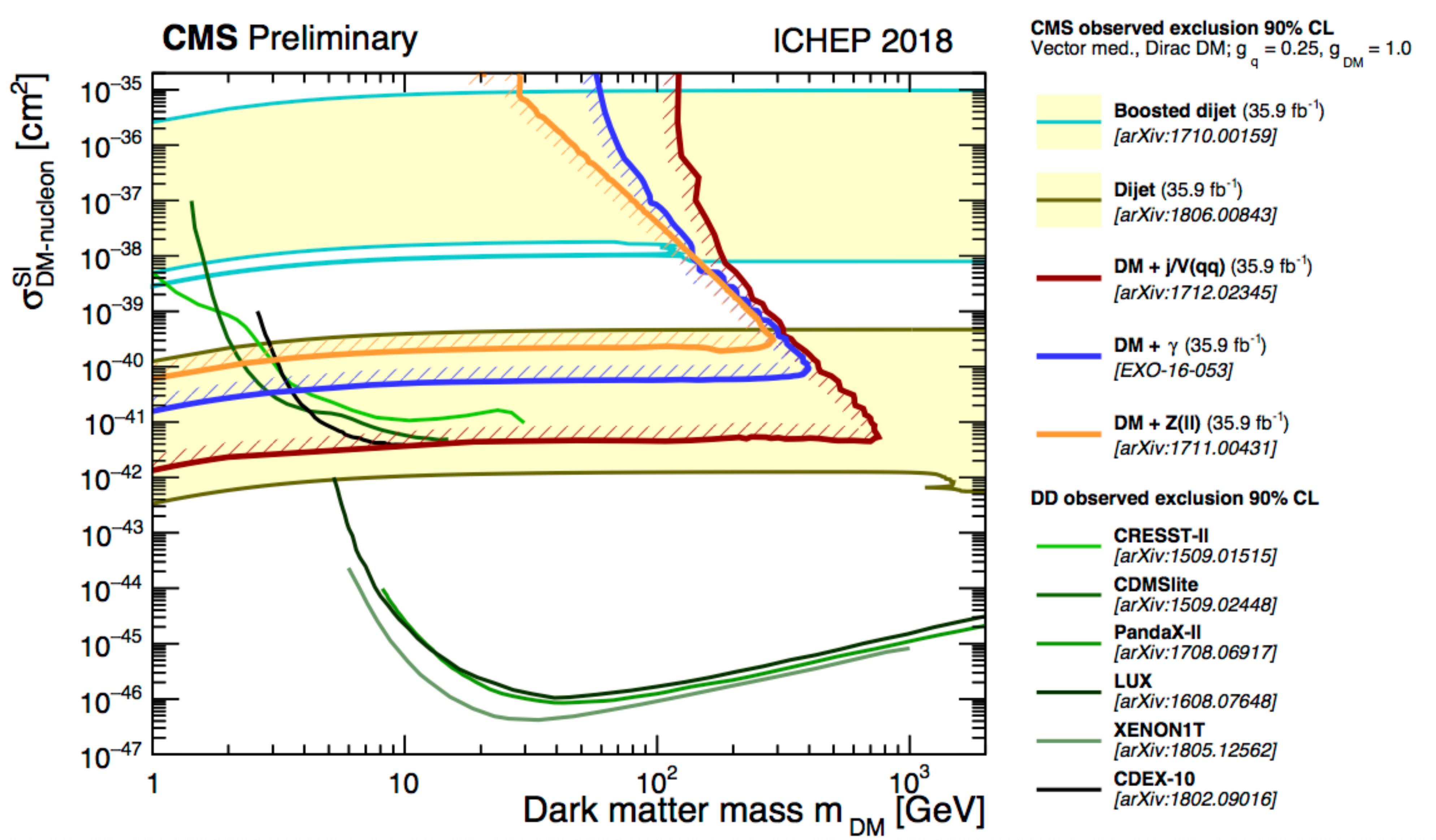}
\end{center}
\caption{CMS DM exclusion limits at 90\% CL in the $m_\text{DM} - \sigma_\text{SI}$ plane for a vector mediator, Dirac DM and couplings $g_q = 0.25$ and $g_\text{DM} = 1.0$ compared with the XENON1T,\cite{Aprile:2018dbl} LUX,\cite{Akerib:2016vxi} PandaX-II,\cite{Cui:2017nnn} CDMSLite\cite{Agnese:2015nto}, CDEX\cite{Jiang:2018pic} and CRESST-II\cite{Angloher:2015ewa} limits, which constitute the strongest constraints in the shown mass range. From Ref.~\protect\refcite{CMSsummary}.}
\label{fg:SI_CMSDD_Summary}
\end{figure}

The complementarity between the results from DM direct detection and related LHC searches is shown in Fig.~\ref{fg:SI_CMSDD_Summary}. CMS limits do not include relic-density constraints and their relative importance strongly depend on the chosen coupling and model scenario. Therefore, the shown CMS exclusion regions in this plot are not applicable to other choices of coupling values or models.

%%%%%%%%%%%%%%%%%%%%%%%%%%%%%%%%%%%%%%%%%%%%%%%%%%%%
\section{Summary}\label{sc:summary}

Signals of dark matter are sought after in direct and indirect detection and in production in colliders. Additional constraints are obtained from cosmological observations on its nature (thermal/non-thermal, super/sub/relativistic, etc.) and on its relic density. Some still unexplained evidence in DD and ID are under investigation taking into account additional systematic uncertainties that matter--DM interactions may hinder and  the strong dependence of the ID \& DD results interpretation on astrophysical assumptions. In parallel, the LHC experiments search for DM in a variety of channels following different  approaches and strategies characterised by high dependency on theoretical models and/or assumptions. Some approaches provide access to the mediator nature than to DM itself, rendering a possible signal difficult to be assigned to DM. The quest for dark matter continues in all fronts.

%%%%%%%%%%%%%%%%%%%%%%%%%%%%%%%%%%%%%%%%%%%%%%%%%%%%
\section*{Acknowledgements}
The author would like to thank the MG15 Meeting organisers for the kind invitation to present this talk. This work was supported by the Generalitat Valenciana via the Project PROMETEO-II/2017/033, by the Spanish MICIU via the grant FPA2015-65652-C4-1-R, by the Severo Ochoa Excellence Centre Project SEV-2014-0398, and by a 2017 Leonardo Grant for Researchers \& Cultural Creators, BBVA Foundation.

%%%%%%%%%%%%%%%%%%%%%%%%%%%%%%%%%%%%%%%%%%%%%%%%%%%%
\bibliography{main}{}
\bibliographystyle{ws-mpla-vaso}

\end{document}